\documentclass[prd,eqsecnum,twocolumn,amsfonts,amssymb,longbibliography]{revtex4-1}

\usepackage{adjustbox}
\usepackage{url}
\usepackage{hyperref} 
\hypersetup{
  colorlinks=true,
        citecolor=,
        linkcolor=,
        filecolor=,
        urlcolor=blue,
}

\usepackage{graphicx}
\usepackage{bm}

\newcommand{\beq}{\begin{equation}}
\newcommand{\eeq}{\end{equation}}
\newcommand{\beqs}{\begin{eqnarray}}
\newcommand{\eeqs}{\end{eqnarray}}
\newcommand{\lsim}{\mathrel{\raisebox{-
.6ex}{$\stackrel{\textstyle<}{\sim}$}}}
\newcommand{\gsim}{\mathrel{\raisebox{-
.6ex}{$\stackrel{\textstyle>}{\sim}$}}}

\begin{document}

\title{Study of the Ultraviolet Behavior of an 
O($N$) $|\vec \phi|^6$ Theory in $d=3$ Dimensions}

\author{Robert Shrock}

\affiliation{C. N. Yang Institute for Theoretical Physics and
Department of Physics and Astronomy, \\
Stony Brook University, Stony Brook, NY 11794, USA }

\begin{abstract}

  We study the ultraviolet (UV) behavior of an O($N$) $|\vec \phi |^6$
  theory in $d=3$ spacetime dimensions, focusing on the question of
  the range in $N$ over which the perturbative beta function exhibits
  robust evidence of a UV zero in the $|\vec \phi |^6$ coupling,
  $g$. The four-loop $(4\ell)$ beta function is known to have a
  (scheme-independent) UV zero at $g=g_{_{UV,4\ell}}$, which is
  reliably calculable for large $N$.  For our analysis we use the
  six-loop beta function calculated in the minimal subtraction
  scheme. We find that this six-loop beta function has a UV zero,
  $g_{_{UV,6\ell}}$, if $N > N_c$, where $N_c \simeq 796$, and we
  calculate $g_{_{UV,6\ell}}$. To investigate the reliability of the
  result in the region of $N \gsim N_c$, we apply three methods: (i)
  calculation of the fractional difference between $g_{_{UV,4\ell}}$
  and $g_{_{UV,6\ell}}$, (ii) a Pad\'e approximant, and (iii) an
  assessment of scheme dependence. Our results provide quantitative
  measures of the range of $N$ over which the six-loop beta function
  has a UV zero and of the $1/N$ corrections to the value of $g$ at
  the UV zero for large but finite $N$. If one imposes a benchmark
  requirement that the fractional difference between $g_{_{UV,4\ell}}$
  and $g_{_{UV,6\ell}}$ must be less than 15 \%, then our results show
  that this requirement is satisfied for $N \gsim 2 \times 10^3$.  The
  possible role of nonperturbative effects is also noted.
  
\end{abstract}

\maketitle


\section{Introduction}
\label{intro_section}

In this paper we study the ultraviolet (UV) behavior of an O($N$)
$|\vec \phi|^6$ quantum field theory in $d=3$ spacetime
dimensions. This theory, commonly denoted $|\vec \phi|^6_3$, involves
an $N$-component real scalar field $\vec \phi = (\phi_1,...,\phi_N)^T$
and is defined by the path integral $Z = \int \prod_x [d \phi_i (x)]
\, e^{iS}$ with $S=\int d^3x \, {\cal L}$, and the bare Lagrangian
\beq
    {\cal L} = \frac{1}{2}(\partial_\nu{\vec \phi}) \cdot
    (\partial^\nu{\vec \phi})
    - \frac{1}{2}m^2 |\vec\phi|^2 
    -\frac{\lambda}{4N}|\vec\phi|^4 -
    \frac{g}{6N^2}|\vec \phi|^6 \ ,
\label{lag}
\eeq
where $|\vec\phi| = (\sum_{i=1}^N \phi_i^2 \, )^{1/2}$.  
In $d=3-\epsilon$ (Euclidean) dimensions, the O($N$) $|\vec\phi|^6$
theory has been extensively analyzed to obtain expressions for critical
exponents describing tricritical points in condensed matter physics
\cite{stephen_mccauley,lewis_adams,lawrie_sarbach,hager}.
Early studies of the theory as a relativistic quantum field theory in
$d=3$ spacetime dimensions include 
\cite{wilson_f63,townsend,pisarski,appelquist_heinz1,appelquist_heinz2,bmb}.

Because of quantum corrections, the physical coupling $g=g(\mu)$
depends on the Euclidean energy/momentum scale, $\mu$, where it is
measured.  This dependence is described by the renormalization group
(RG) beta function \cite{callan,symanzik,wilson_rg} of the theory,
$\beta_g = dg/d\ln\mu$.  The lowest-order (two-loop, $O(g^2)$) term in
$\beta_g$ is positive \cite{stephen_mccauley}, so this theory is
infrared (IR)-free, i.e., $g(\mu) \to 0$ as $\mu \to 0$.  An important
question is whether, for a given $N$, the theory has a UV zero in
$\beta_g$ at some value $g_{_{UV}}$. If this is the case and if a
perturbative analysis is adequate to describe the physics, then, as
the reference energy/momentum scale $\mu$ increases from 0 to
$\infty$, $g(\mu)$ increases from 0 and approaches $g_{_{UV}}$ from
below.  Since the coefficients of the quadratic and quartic terms in
the Lagrangian (\ref{lag}) are both dimensionful, and since $\lim_{\mu
  \to \infty} m^2/\mu^2 = 0$ and $\lim_{\mu \to \infty} \lambda/\mu =
0$, they are expected to play a negligible role in the ultraviolet
limit $\mu \to \infty$. We denote the UV zero (presuming that it
exists) of the $n$-loop ($n\ell$) beta function as $g_{_{UV,n\ell}}$.
The term of order $g^p$ in $\beta_g$ arises from graphs with a maximum
number of loops $n$ given by $n=2(p-1)$. The $O(g^3)$ term in the beta
function is negative, so that at this order, this four-loop ($4\ell$)
beta function, $\beta_{g,4\ell}$, has a UV zero
\cite{townsend,pisarski,appelquist_heinz2}. In the large-$N$ limit,
with the normalization in Eq. (\ref{lag}), this occurs at the value of
the coupling $g_{_{UV,4\ell}}=192$. It was noted in
\cite{townsend,pisarski} that the $N$-dependence of higher-loop terms
in $\beta_g$ is such that for large $N$, the inclusion of these
higher-loop terms would produce only a small fractional shift $\propto
1/N$ in the value of the coupling at the UV zero, and therefore the
calculation of the value should be reliable in the large-$N$ limit.
Such a UV zero in the beta function is a UV fixed point (UVFP) of the
renormalization group.  The existence of a UVFP in an IR-free theory
is of considerable interest, since it means that one has perturbative
control of the theory in both the IR and UV limits.  A previous
example of an IR-free theory with a UVFP is the nonlinear $O(N)$
$\sigma$ model in $d=2+\epsilon$ dimensions
\cite{brezin_zinnjustin_prl,brezin_zinnjustin,nlsm,polyakov}.  The
early studies \cite{townsend,pisarski} also cautioned that at small
and moderate $N$, this formal UV zero at $g=g_{_{UV,4\ell}}$ might be
an artifact of the perturbative calculation.

We briefly review some further relevant work on this theory. After the
studies in
\cite{townsend,pisarski,appelquist_heinz1,appelquist_heinz2}, a
variational calculation was carried out in the $N \to \infty$ limit by
Bardeen, Moshe, and Bander (BMB) in \cite{bmb}, who found that for $g
> g_{cr}$, where $g_{cr}=(4\pi)^2 \simeq 158$, the theory undergoes a
transition to a strongly coupled phase involving dynamical mass
generation for the scalar field and spontaneous breaking of scale
invariance, with the resultant appearance of a massless
Nambu-Goldstone boson (NGB), namely a dilaton \cite{bmb}.  Since
$g_{cr} < g_{_{UV,4\ell}}$, it was concluded in \cite{bmb} that in the
large-$N$ limit where the BMB calculations were performed, the physics
is described by the properties of this strong-coupling phase rather
than by a UV zero in the (perturbative) beta function.  The properties
in the $N \to \infty$ limit were further studied in
\cite{dkn1,dkn2,amit_eliezer}.  Exploratory lattice studies to probe
the BMB phase were performed in \cite{kn_lattice,karsch}. Refs.
\cite{gdottir1,gdottir2} argued that at finite $N$, the BMB phase is
unstable.  More recently, Ref.  \cite{osw} investigated the effect of
higher-order corrections in $1/N$ on the BMB dilaton and found that it
becomes a tachyon when one takes account of these $1/N$ corrections.
On this basis, the authors of Ref. \cite{osw} concluded that at finite
$N$, the BMB phase with spontaneously broken approximate scale
invariance is unstable.  Some recent related studies of this theory
include \cite{litim_marchais_mati,litim_trott,yd1,yd2,yd3}.

In parallel with these continuing studies of the role of possible
nonperturbative effects at finite $N$ in the $|\vec\phi|^6_3$ theory,
it seems worthwhile to investigate the UV properties of the
(perturbative) beta function further. At the time of the early studies
in \cite{townsend,pisarski,appelquist_heinz1,appelquist_heinz2,bmb},
$\beta_g$ had been calculated only up to $O(g^3)$.  Subsequently, it
was calculated to $O(g^4)$ in \cite{hager}, and this remains the
highest order to which it has been computed.  A very basic question
that, to our knowledge, has not been studied yet, is whether, for a
given $N$, this $O(g^4)$ beta function exhibits evidence for a
(reliably calculable) UV zero, denoted $g_{_{UV,6\ell}}$. We address
this question in this paper.

A necessary condition for such evidence is that the values of the
coupling at this UV zero obtained from calculations of the beta
function to successive orders in $g$ should be close to each other.
To determine the region in $N$ over which this condition is satisfied,
we will compare the values of $g_{_{UV,4\ell}}$ and $g_{_{UV,6\ell}}$
(for $N$ values where the latter exists).  Furthermore, the terms in
the beta function at order $g^p$ with $p \ge 4$ depend on the scheme
used for regularization and renormalization.  By itself, this property
does not render these higher-order terms unphysical; for example,
higher-order calculations of quark and gluon scattering in quantum
chromodynamics (QCD) are also scheme-dependent, but still play a
crucial role in the analysis of experimental data, and work continues
on the construction and application of optimal schemes for QCD
calculations (see, e.g., \cite{brodsky} and references therein).
However, this does mean that one must assess the effect of this
scheme-dependence, and we shall do this as part of our study. In
carrying out this study, it should be stressed that nonperturbative
effects may be important, and we refer the reader to the continuing
analysis of this topic, e.g. in \cite{osw,yd3}, as well as earlier
works including \cite{bmb}.  However, bearing this caveat in mind, one
should at least elucidate the predictions from the beta function
calculated to the highest order to which it is known for general $N$,
and that is the purpose of our present study.

In passing, it should be mentioned that theories of $\phi^6_3$ type
with various global symmetries, representations of the scalar fields, and
sixth-degree interaction terms have also been of recent interest in the
context of large-charge expansions and conformal field theory, e.g.,
\cite{hellerman_largecharge,alvarez-gaume_largecharge,cuomo_largecharge,sannino_largecharge,jack_jones2020}.
Here we will confine our analysis to the simple realization of this
theory in the Lagrangian (\ref{lag}).

This paper is organized as follows. In Section \ref{beta_section} we
present the results of an analysis of the evidence, for a given $N$, of a
UV zero the six-loop beta function. In Section \ref{pade_section} we
apply the method of Pad\'e approximants to study this question
further.  Section \ref{scheme_section} contains an assessment of the
effects of scheme dependence. Our conclusions are presented in Section
\ref{conclusion_section}. Some auxiliary relations are given
in an Appendix. 


\section{Beta Function and UV Zero}
\label{beta_section}

In this section we analyze the beta function of the $O(N)$
$|\vec\phi|^6_3$ theory.  The beta function $\beta_g = dg/d\ln\mu$ has a
series expansion in powers of the interaction coupling $g$, starting
with a term of $O(g^2)$,
\beq
\beta_g = g\sum_{j=1}^\infty  b_j g^j \ . 
\label{betaseries}
\eeq
In the literature there are several different normalization
conventions for the $|\vec\phi|^6$ coupling; for the reader's
convenience, in the Appendix we list conversion formulas relating some
of these.  As noted above, the term in $\beta_g$ of $O(g^p)$ arises
from graphs with a maximal number of loops equal to $n=2(p-1)$.  We
denote the truncation of the infinite series (\ref{betaseries}) to
$O(g^p)$, as $\beta_{g,n\ell}$, where $n\ell$ is short for
$n$-loops. As is the case with other scalar field theories, although
Eq. (\ref{betaseries}) is an asymptotic expansion
\cite{lipatov,bgz76}, it can still yield useful information
about the properties of the theory.

With the normalization in Eq. (\ref{lag}), the first two
coefficients, $b_j$, $j=1,2$, are
\cite{townsend,pisarski,appelquist_heinz2}
(see also \cite{stephen_mccauley,lewis_adams}) 
\beq
b_1 = \frac{3N+22}{2\pi^2 N^2} 
\label{b1}
\eeq
and
\begin{widetext}
\beqs
b_2 &=& - \frac{1}{2^7 \pi^2 N^4} \bigg [ (N^3+34N^2+620N+2720)
  +\frac{8}{\pi^2} (53N^2+858N+3304) \bigg ] \cr\cr
&=& \frac{1}{N^4}\bigg [(0.791572 \times 10^{-3})N^3 + 0.0609195N^2+1.04129N+4.27300 \bigg ] \ , 
\label{b2}
\eeqs
\end{widetext}
where floating-point numbers are given to the indicated accuracy.  As
mentioned before, these first two terms of $O(g^2)$ and $O(g^3)$ in
$\beta_g$ contain the maximal scheme-independent information in this
function. The UV zero in the four-loop beta function occurs at
$g_{_{UV,4\ell}} = -b_1/b_2$, namely
\begin{widetext}
\beq
g_{_{UV,4\ell}} = \frac{64N^2(3N+22)}{(N^3+34N^2+620N+2720)+
  (8/\pi^2)(53N^2+858N+3304)} \ .
\label{guv_4loop}
\eeq
This is a monotonically decreasing function of $N$. For large $N$,
\beqs
g_{_{UV,4\ell}} &=& 192 \bigg [ 1 - \frac{8(159+10\pi^2)}{3\pi^2 N} 
  +\frac{4(134832+14144\pi^2+215\pi^4)}{3\pi^4 N} + 
  O\Big ( \frac{1}{N^3} \Big ) \bigg ] \cr\cr
&=& 192 \bigg [ 1 - \frac{69.62685}{N} + \frac{4043.0263}{N^2} + 
  O\Big ( \frac{1}{N^3} \Big )  \bigg ] \ . 
\label{guv_4loop_largeN}
\eeqs

The coefficient $b_3$ of the $g^4$ term in $\beta_g$ has
been calculated in \cite{hager} in the minimal subtraction scheme
\cite{hooft72,msbar}. With the normalization in Eq. (\ref{lag}), it is  
\beqs
b_3 &=& \frac{1}{2^7\pi^6 N^6} 
\Big ( 1857N^3+45976N^2+367716N+950576 \Big )  + 
\frac{1}{2^9 \pi^4N^6}\Big ( 36N^4+1607N^3+33568N^2+273772N+735392 \Big )
\cr\cr
&-& \frac{3 \ln(2)}{2^8 \pi^4 N^6}\Big ( N^4+N^3-700N^2-8236N-24816 \Big ) 
+ \frac{5}{2^{11}\pi^2 N^6}\Big ( N^4+64N^3+1352N^2+12248N+36960 \Big ) \cr\cr
&-& \frac{105\zeta(3)}{2^8 \pi^6 N^6}\Big (11N^3+428N^2+4228N+12208 \Big ) + 
\frac{1}{2^6 \pi^6 N^6}
\Big ( 24\beta(4)+\pi^2 G \Big )\Big (31N^3+1126N^2+11876N+37592 \Big ) \ ,
\cr\cr
 &=& \frac{1}{N^6} \Big [ (0.885804 \times 10^{-3})N^4 +
  0.0739318 N^3 + 1.81979 N^2 + 16.3518 N + 47.4455 \, \Big ]  \ ,
\label{b3}
\eeqs
\end{widetext}
where the Dirichlet beta function $\beta(s)$ is defined as
\beq
\beta(s) = \sum_{n=0}^\infty \frac{(-1)^n}{(2n+1)^s} \ , 
\label{dirichlet_beta}
\eeq
and $G=\beta(2)$ is the Catalan constant, with the values
$G=0.915965594$ and $\beta(4)=0.98894455$ to the given floating-point
accuracy.  As is evident from the numerical evaluation of $b_3$ in
Eq. (\ref{b3}), it is positive for all physical $N$ and is a
monotonically increasing function of $N$.  With the normalization
convention in Eq. (\ref{lag}), these coefficients have the large-$N$
behavior $b_1 \sim 1/N$, $b_2 \sim 1/N$, and $b_3 \sim 1/N^2$.

We now address the question of whether, for a given $N$, the six-loop
beta function, $\beta_{g,6\ell}$, with $b_3$ computed in the minimal
subtraction scheme, has a (perturbatively reliably calculable) UV
zero.  Aside from the double IR-zero at $g=0$, the zeros of the
$n$-loop beta function $\beta_{g,n\ell}$, i.e., the beta function
calculated to $O(g^p)$, where $p=(n/2)+1$, are the solutions of the
algebraic equation $\sum_{j=1}^{n/2} b_j g^{j-1}=0$. For
$\beta_{g,6\ell}$, this is the equation $b_1+b_2g + b_3 g^2=0$, with
solutions $g_\pm = (2b_3)^{-1}(-b_2 \pm \sqrt{b_2^2 - 4b_1b_3} \, )$.
With $N$ formally generalized from positive integers to positive real
numbers, $b_2^2-4b_1b_3$ is positive if $N > N_c = 796.111$ and
negative for the rest of the physical range $1 \le N < N_c$.  Hence,
for $N > N_c$, the six-loop beta function $\beta_{g,6\ell}$ has two
zeros on the positive $g$ axis, and the one nearer to the origin is
the UV zero, $g_{_{UV,6\ell}} = g_- =
(2b_3)^{-1}(-b_2-\sqrt{b_2^2-4b_1b_3} \, )$.  This has the large-$N$
series expansion
\begin{widetext}
  \beqs
g_{_{UV,6\ell}}
&=& 192 \bigg [ 1 + \frac{4(978+25\pi^2-216\ln 2)}{3\pi^2 N} +
  \frac{4}{3\pi^4N^2} \Big ( 3232704 +212612 \pi^2 +4565 \pi^4 +8928 \pi^2 G
  -1218249 \ln 2 \cr\cr
  &+& 124416 (\ln 2)^2 -33192 \pi^2 \ln 2 - 83160 \zeta(3) \Big )
   +O\Big ( \frac{1}{N^3} \Big ) \bigg ] \cr\cr
&=& 192 \bigg [ 1 + \frac{145.230}{N} + \frac{67847.343}{N^2} + 
    O\Big ( \frac{1}{N^3} \Big ) \bigg ] \ .
\label{guv_6loop_largeN}
\eeqs
\end{widetext}

It follows that
\beq
\lim_{N \to \infty} g_{_{UV,6\ell}} =
\lim_{N \to \infty} g_{_{UV,4\ell}} = 192 \ . 
\label{guv_64ell}
\eeq
Note that the $1/N$ correction to $g_{_{UV,4\ell}}$ is negative, while
the $1/N$ correction to $g_{_{UV,6\ell}}$ is positive.
Given the $N$-dependence of still higher-order terms $O(g^p)$ with $p
\ge 5$ in $\beta_g$, as discussed in \cite{pisarski}, the result
(\ref{guv_64ell}) can be generalized to
\beq
\lim_{N \to \infty} g_{_{UV,n\ell}} =
\lim_{N \to \infty} g_{_{UV,4\ell}} = 192 \ , 
\label{guv_n4ell}
\eeq
where $n=2(p-1)$ with $p \ge 4$. 

A necessary condition for a credible zero of a beta function is that
when one calculates it to successive orders of perturbation theory,
one obtains values that are close to each other, i.e., values with small
fractional differences. We define the fractional difference as
\beq
\Delta(g_{_{UV,n\ell}},g_{_{UV,n'\ell}}) \equiv
\frac{g_{_{UV,n'\ell}}-
     g_{_{UV,n\ell}}}{g_{_{UV,n\ell}}} \ .
  \label{delta}
  \eeq
In the large-$N$
limit, the fractional difference between $g_{_{UV,4\ell}}$ and
$g_{_{UV,6\ell}}$ is
\begin{widetext}
  \beqs
  \Delta(g_{_{UV,4\ell}},g_{_{UV,6\ell}} ) 
  &=& \frac{12(144+5\pi^2-24\ln 2)}{\pi^2 N} +
  \frac{4}{\pi^4 N^2} \Big ( 1215792 + 84036\pi^2 + 1850\pi^4 + 2976\pi^2 G
    -436608\ln 2 \cr\cr
    &+& 41472(\ln 2)^2 -12984\pi^2\ln 2 - 27729\zeta(3) +
    71424\beta(4) \Big ) + O \Big ( \frac{1}{N^3} \Big ) \cr\cr
  &=& 
  \frac{214.8566}{N} + \frac{78764.106}{N^2} +   O \Big ( \frac{1}{N^3} \Big ) 
\ . 
  \label{Delta_4loop_6loop}
  \eeqs
  \end{widetext}

  This fractional difference vanishes as $N \to \infty$, in agreement
  with the conclusion reached in
  \cite{townsend,pisarski,appelquist_heinz2} (before $b_3$ had been
  calculated). The new information obtained here is the calculation of
  the series expansion in powers of $1/N$ in
  Eq. (\ref{Delta_4loop_6loop}), which provides a quantitative measure
  of the accuracy and reliability of the perturbative calculation of
  the value of the coupling at the UV zero for a given large $N$.  In
  Table \ref{guv_table} we list $g_{_{UV,4\ell}}$, $g_{_{UV,6\ell}}$,
  and the fractional difference
  $\Delta(g_{_{UV,4\ell}},g_{_{UV,6\ell}})$ for some illustrative
  values of $N$. As $N$ increases well beyond $N_c$, this fractional
  difference decreases reasonably quickly.  For example, for $N=2
  \times 10^3$, $N=4 \times 10^3$, and $N=10^4$,
  $\Delta(g_{_{UV,4\ell}},g_{_{UV,6\ell}})$ has the approximate values
  13 \%, 6 \%, and 2 \%, respectively. Thus, if one imposes  
  a requirement that the fractional difference
  $\Delta(g_{_{UV,4\ell}},g_{_{UV,6\ell}})$ must be less than, say, 15 \%,
  in order for the calculation of the value of the UV zero to be
  reasonably reliable, then our results show that this criterion is
  satisfied for $N \gsim 2 \times 10^3$. 


\section{Analysis With Pad\'e Approximants}
\label{pade_section}

One can gain further insight into the behavior of the beta function by
the use of Pad\'e approximants.  Given a series expansion $f(z) =
\sum_{n=0}^{n_{\rm max}} a_n z^n$, the $[r,s]$ Pad\'e approximant (PA)
is the rational function with numerator and denominator polynomials in
$z$ of degree $r$ and $s$, respectively, where $r+s=n_{\rm max}$, such
that the Taylor series expansion of this rational function matches the
series expansion for $f(z)$ to its highest order, $n_{\rm max}$.  The
Pad\'e method can be considered to be semi-perturbative, since it uses
as input a perturbative series expansion but produces a closed-form
rational function, whose higher-order terms of order $z^n$ with $n >
n_{\rm max}$ are thus determined. Since the double IR zero in
$\beta_g$ at $g=0$ is not relevant here, it will be convenient to
consider the reduced (red.)  beta function normalized so that it is
equal to 1 for $g=0$:
\beq
\beta_{g,{\rm red.}} \equiv \frac{\beta_g}{\beta_{g,2\ell}} =
\frac{\beta_g}{b_1 g^2} =
1 + \frac{1}{b_1} \sum_{j=2}^\infty b_j g^{j-1} \ .
\label{betareduced}
\eeq
From the beta function calculated to $O(g^p)$, 
one thus obtains the reduced beta function of degree
$(p-2)$ in $g$.  In particular, from $\beta_{g,6\ell}$, we
have $\beta_{g,{\rm red.},6\ell} = 1+(b_2/b1)g+(b_3/b_1)g^2$. We
denote the Pad\'e approximants to $\beta_{g,{\rm red.},n\ell}$
simply as $[r,s]_{n\ell}$. The PA $[2,0]_{6\ell}$ is this function
itself, which we have already analyzed, and the PA $[0,2]_{6\ell}$ has
no zero, so we study the $[1,1]_{6\ell}$ Pad\'e approximant. In terms
of the coefficients $b_j$, $j=1,2,3$, this is
\beq
    [1,1]_{6\ell} = \frac{1+\Big ( \frac{b_2^2-b_1b_3}{b_1b_2} \Big )g}
    {1-\Big (\frac{b_3}{b_2}\Big )g} \ .
\label{p11}
\eeq
We label the zero of this $[1,1]_{6\ell}$ PA as
$g_{_{UV,[1,1]_{6\ell}}}$. This is
\beq
g_{_{UV,[1,1]_{6\ell}}} = \frac{b_1b_2}{b_1b_3-b_2^2} =
\frac{-\Big ( \frac{b_1}{b_2} \Big )}
{1 - \Big ( \frac{b_1 b_3}{b_2^2} \Big )} \ .
\label{guvp11}
\eeq

We list illustrative values of $g_{_{UV,[1,1]_{6\ell}}}$ in Table
\ref{guv_table}.  As is evident from the last term in
Eq. (\ref{guvp11}), $g_{_{UV,[1,1]_{6\ell}}}$ is related to the
value of the UV zero of the four-loop beta function,
$g_{_{UV,4\ell}}=-b_1/b_2$, via division by the factor $1-(b_1
b_3/b_2^2)$. Now $(b_1 b_3/b_2^2) > 0$, so if $(b_1 b_3/b_2^2) < 1$,
then $g_{_{UV,[1,1]_{6\ell}}} > g_{_{UV,4\ell}}$.  Since
$b_1 b_3/b_2^2 \sim O(N^{-1})$ as $N \to \infty$, it follows that
\beq
\lim_{N \to \infty} g_{_{UV,[1,1]_{6\ell}}} =
\lim_{N \to \infty} g_{_{UV,4\ell}} \ .
\label{p114ell}
\eeq
For $N \gg 1$, $g_{_{UV,[1,1]_{6\ell}}}$ has the expansion
\begin{widetext}
\beqs
g_{_{UV,[1,1]_{6\ell}}} &=& 192 \Big [ 1 +
  \frac{4(978+25\pi^2-216\ln 2)}{3\pi^2 N} +
  \frac{4}{3\pi^4N^2}\Big (993216+57092\pi^2+1865\pi^4+8928\pi^2G
  -471744\ln 2 \cr\cr
    &+& 62208(\ln 2)^2 -83160\zeta(3) +214272\beta(4) \, \Big )
+ O\Big ( \frac{1}{N^3} \Big ) \Big ] \cr\cr
&=& 192 \Big [ 1 + \frac{145.230}{N} + \frac{21683.974}{N^2}
  + O\Big ( \frac{1}{N^3} \Big ) \Big ] \ . 
\label{guvp11_largeN}
\eeqs
As is evident from Eqs. (\ref{guv_6loop_largeN}) and (\ref{guvp11_largeN}),
$g_{_{UV,6\ell}}$ and $g_{_{UV,[1,1]_{6\ell}}}$ have the same leading
$1/N$ correction terms.  This can be understood as a consequence of the fact
that the $[1,1]_{6\ell}$
Pad\'e approximant incorporates information from the $b_3 g^4$
term in $\beta_{g,6\ell}$, or equivalently, the $(b_3/b_1)g^2$ term
in $\beta_{g,{\rm red.},6\ell}$.

For large $N$, the fractional differences
$\Delta(g_{_{UV,4\ell}},g_{_{UV,[1,1]_{6\ell}}})$ and
$\Delta(g_{_{UV,4\ell}},g_{_{UV,[1,1]_{6\ell}}})$ are
\beqs
\Delta(g_{_{UV,4\ell}},g_{_{UV,[1,1]_{6\ell}}}) &=&
\frac{12(144+5\pi^2-24\ln 2)}{\pi^2N} +
\frac{4}{\pi^4N^2}\Big ( 469296+32196\pi^2+950\pi^4+2976\pi^2G
-187776\ln 2 \cr\cr
&+&20736(\ln 2)^2 -4344\pi^2\ln 2 -27729\zeta(3)+71424\beta(4) \Big )
+ O\Big ( \frac{1}{N^3} \Big ) \cr\cr
&=& \frac{214.857}{N} + \frac{32600.74}{N^2} +
O\Big ( \frac{1}{N^3} \Big )  
\label{Delta_guv4loop_gp11_largeN}
\eeqs
\end{widetext}
and
\beqs
\Delta(g_{_{UV,6\ell}},g_{_{UV,[1,1]_{6\ell}}}) &=&
-\frac{144(144+5\pi^2-24\ln 2)^2}{\pi^4N^2} 
+ O\Big ( \frac{1}{N^3} \Big ) \cr\cr
&=& -\frac{46163.369}{N^2} + O\Big ( \frac{1}{N^3} \Big ) \ .
\label{Delta_guv6loop_gp11_largeN}
\eeqs
Since $g_{_{UV,6\ell}}$ and $g_{_{UV,[1,1]_{6\ell}}}$ have the same
$1/N$ correction terms in the large-$N$ limit, as observed above, it
follows that $\Delta(g_{_{UV,6\ell}},g_{_{UV,[1,1]_{6\ell}}})$ vanishes
like $1/N^2$ in this limit.  This is in contrast to 
$\Delta(g_{_{UV,4\ell}},g_{_{UV,{6\ell}}})$ and 
$\Delta(g_{_{UV,4\ell}},g_{_{UV,[1,1]_{6\ell}}})$, which both
vanish like $1/N$ for large $N$. 

In order for $g_{_{UV,[1,1]_{6\ell}}}$ to be
acceptable as a UV zero, it is necessary that there should not be
a pole in the $[1,1]_{6\ell}$ PA on the positive real axis closer
to the origin. The pole in this approximant occurs at
\beq
g_{[1,1]_{6\ell},{\rm pole}} = \frac{b_2}{b_3} \ .
\label{wp11_pole}
\eeq
Since $b_2 < 0$ and $b_3 > 0$ for all physical $N$, this pole occurs
on the negative real axis, thereby fulfilling the above necessary condition.
As $N \to \infty$, the value of $g$ at this pole behaves as
\beqs
g_{[1,1]_{6\ell},{\rm pole}} &=& -\frac{16\pi^2N}{144+5\pi^2-24\ln 2}
+ O(1) \cr\cr
&=& -0.8936N + O(1) \ . 
\label{gp11pole_largeN}
\eeqs

In general, in the large-$N$ limit, the $1/N$ expansions given above show
that 
\beq
g_{_{UV,4\ell}} \le g_{_{UV,[1,1]_{6\ell}}} \le g_{_{UV,6\ell}} \ .
\label{grel}
\eeq
and
\beq
|\Delta(g_{_{UV,6\ell}},g_{_{UV,[1,1]_{6\ell}}})| \le
\Delta(g_{_{UV,4\ell}}, g_{_{UV,[1,1]_{6\ell}}})
\ . 
\label{deldelrel}
\eeq
with equality at $N = \infty$. 

As $N$ decreases from large values, $b_1 b_3/b_2^2$ increases, and as
$N$ decreases below a value $N_d \simeq 150.799$, this ratio increases
through 1, producing a pole in $g_{_{UV,[1,1]_{6\ell}}}$.  Clearly,
this method of obtaining an estimate of a UV zero in $\beta_{g,6\ell}$
via a zero in the $[1,1]_{6\ell}$ approximant at $g_{_{UV,[1,1]_{6\ell}}}$ is
only reliable for values of $N$ well above $N_d$.

  Thus, the $[1,1]_{6\ell}$ Pad\'e approximant to $\beta_{g,{\rm
      red.},6\ell}$ yields a UV zero over a larger range of $N$ than
  the beta function itself, extending below $N_c \simeq 796$ to the
  vicinity of $N_d \simeq 151$. However, as noted above, as $N$
  approaches the vicinity of $N_d$ from above, the value of
  $g_{_{UV,[1,1]_{6\ell}}}$ deviates substantially from the
  scheme-independent value, $g_{_{UV,4\ell}}$.  For example, at an
  illustrative value below $N_c$ but above $N_d$, namely $N=300$,
  although the [1,1] Pad\'e approximant has a UV zero,
  $g_{_{UV,[1,1]_{6\ell}}}=378.17$, this is not close to
  $g_{_{UV,4\ell}}=154.71$.  Consequently, in this vicinity, the
  method does not satisfy the requirement that different perturbative
  or semi-perturbative methods of calculating this UV zero should
  yield values in approximate agreement with each other.  Among the
  entries in Table \ref{guv_table}, in addition to the values of
  $g_{_{UV,[1,1]_{6\ell}}}$ themselves, we list the fractional
  difference between $g_{_{UV,4\ell}}$ and $g_{_{UV,[1,1]_{6\ell}}}$,
  denoted $\Delta(g_{_{UV,4\ell}},g_{_{UV,[1,1]_{6\ell}}})$ and the
  fractional difference between $g_{_{UV,6\ell}}$ and
  $g_{_{UV,[1,1]_{6\ell}}}$, denoted
  $\Delta(g_{_{UV,6\ell}},g_{_{UV,[1,1]_{6\ell}}})$.
  

\section{Assessment of Scheme Transformations}
\label{scheme_section}

Since $b_3$ and $g_{_{UV,6\ell}}$ are scheme-dependent, one
should assess the effect of this scheme dependence in a study of a UV zero
of the beta function for this theory.  A scheme transformation can be
expressed as a mapping between $g$ and
$g'$, which we write as $g = g' f(g')$, where 
$f(g')$ is the scheme transformation function, satisfying $f(0)=1$. 
We will consider functions $f(g')$ that have a Taylor series expansion 
\beq
f(g') = 1 + \sum_{j=1}^{j_{\rm,max}} k_j \, g'^j \ , 
\label{fwprime}
\eeq
where the $k_s$ are constants, where $j_{\rm max}$ may be finite or
infinite. The Jacobian of the transformation is
$J=dg/dg'$, 
\beq
J = 1 + \sum_{j=1}^{j_{max}} (j+1)k_j g'^j \ . 
\label{j}
\eeq
After the scheme transformation is applied, the beta function in the
new scheme has the form (\ref{betaseries}) with $g$ replaced by
$g'$ and $b_j$
replaced by $b'_j$.  Expressions for the $b_j'$ in terms of the $b_j$
and $k_s$ were derived in \cite{scc,sch}.  Aside from $b_1'=b_1$ and
$b_2'=b_2$, these relations include
\beq
b_3' = b_3 + k_1b_2+(k_1^2-k_2)b_1 \ 
\label{b3prime}
\eeq
\beq
b_4' = b_4 + 2k_1b_3+k_1^2b_2+(-2k_1^3+4k_1k_2-2k_3)b_1 \ ,
\label{b4prime}
\eeq
and so forth for $b_j'$ with higher $j$.

As was discussed in \cite{scc,sch} and studied further in
\cite{sch2,sch3,schl,schi}, in order to be physically
acceptable, a scheme transformation must satisfy several necessary
conditions, which were denoted C$_1$ to C$_4$.  The first two
conditions, C$_1$ and C$_2$, are that the scheme transformation must
map a real positive $g$ to a real positive $g'$ and should not map a
moderate value of $g$, for which perturbation theory may be reliable,
to a value of $g'$ that is so large that perturbation theory is
unreliable. The third condition, C$_3$, is that the Jacobian should
not vanish in the region of $g$ and $g'$ of interest or else the
transformation would be singular. A fourth condition given in
\cite{scc,sch} is that, since the existence of a UV zero of the beta
function is a physical property, a scheme transformation should
satisfy the property that $\beta_g$ has an UV zero if and only if
$\beta_{g'}$ has an UV zero.  These conditions can easily be satisfied
by scheme transformations applied in the vicinity of a zero of the
beta function at sufficiently small coupling, but they are not
automatically satisfied, and are a significant restriction, on a
scheme transformation applied in the vicinity of a generic zero of the
beta function away from the origin.  These results on scheme
transformations have been applied to the study of an IR zero in the
beta function of an asymptotically free theory, such as a non-Abelian
gauge theory with a certain content of massless Dirac fermions in
$d=4$ dimensions \cite{sch,sch2,sch3,tr1,tr2,gracey_simms,schl,schi}
and the Gross-Neveu model \cite{gn} in $d=2$ \cite{bgn}. They have
also been applied to assess scheme dependence in probing for a
possible UV zero in the beta function of an IR-free theory such as
$O(N)$ $|\vec\phi|^4$ theory in $d=4$ \cite{lam,lam3} (reviews include
\cite{zinnjustin,kleinert}).  

Since $\beta_{g,6\ell}$ does not have a UV zero if $N < N_c$,
whereas $\beta_{g,4\ell}$ has, at least formally, a UV zero for all
physical $N$, a natural method to use to study the effects of scheme
dependence is to construct a scheme transformation that eliminates the
$O(g^4)$ term in $\beta_{g',6\ell}$ and thus yields a beta function
consisting of just the first two (scheme-independent) terms.  Since
the beta function in the transformed scheme always has a UV zero, this
scheme transformation would not satisfy condition C$_4$. However, by
applying it, one can at least gain some information about the degree
of scheme dependence in the evidence for or against the property that,
at a given $N \lsim N_c$, the six-loop beta function in a particular scheme
has a UV zero.

To carry out this procedure, we will use the results of
Refs. \cite{sch,sch2,sch3}, which presented scheme transformations
that can be used to set $b_\ell'=0$ for $\ell \ge 3$, thereby reducing
the beta function to its two scheme-independent terms (the 't Hooft
scheme). The simplest way to do this is to set $k_1=0$ in
Eq. (\ref{fwprime}) and then solve the equation $b_3'=0$ for $k_2$,
obtaining
\beq
k_2 = \frac{b_3}{b_1} \ .
\label{k2sol}
\eeq
Although we will only need to apply the procedure here to set
$b_3'=0$, since this is the highest-order coefficient that has been
calculated for this theory, we briefly review how the procedure works
if one has a beta function calculated to higher order.  One next
substitutes the value of $k_2$ from Eq. (\ref{k2sol}) into the
equation for $b_4'$, Eq.  (\ref{b4prime}), and solves the equation
$b_4'=0$ for $k_3$. This procedure is applied iteratively to solve for
$k_j$ with $j \ge 4$ so as to render $b_{j+1}'=0$.  At least formally,
a solution is guaranteed, since the condition that $b_{j+1}'=0$ is a
linear equation for $k_j$ for all $j \ge 2$.  However, while this
procedure can be carried out for sufficiently weak coupling, e.g., in
the deep UV limit of a UV-free theory such as QCD, as originally noted
by 't Hooft \cite{hooft_scheme}, Refs. \cite{sch,sch2,sch3} showed
that it can be more difficult to do this with a physically acceptable
scheme transformation when studying a zero of the beta function away
from the origin in coupling constant space (an IR zero in a UV-free
theory or a UV zero in an IR-free theory).

Given that we take $k_1=0$ in $f(g')$, the procedure for constructing
and applying this scheme transformation here requires only the
determination of a single parameter, $k_2$, since we only have to
eliminate $b_3'$ to reduce the six-loop beta function to its minimal
scheme-independent first two terms in the transformed scheme.  Thus,
we construct a scheme transformation with $j_{\rm max}=2$, $k_1=0$ and
$k_2=b_3/b_1$, as in Eq. (\ref{k2sol}), so as to render $b_3'=0$.
This is the transformation with $f(g')=1+(b_3/b_1)g'^2$,
namely 
\beq
g = g'\Big [ 1+ \Big ( \frac{b_3}{b_1} \Big )g'^2 \Big ] \ .
\label{schemetrans}
\eeq
Then, since $b_3'=0$, the beta function in the transformed scheme is 
\beq
\beta_{g',6\ell} = b_1 g'^2 + b_2 g'^3 \ .
\label{betaprime_3loop}
\eeq
Note that $b_3/b_1 \sim 1/N$ for large $N$, so as $N \to \infty$, the
scheme transformation (\ref{schemetrans}) approaches the identity
mapping.

To check whether, for a given $N$, this scheme transformation
satisfies at least the first three conditions for acceptibility, one
then calculates how close $g$ and the corresponding $g'$ are to each
other.  For a given $N$ and resultant UV zero of the four-loop beta
function, $g_{_{UV,4\ell}}$, one thus solves the cubic equation
(\ref{schemetrans}) for $g'$ with $g= g_{_{UV,4\ell}}$, using the
minimal positive real root as $g'$.  In Table \ref{gprime_table} we
list illustrative values of $N$ and $g_{_{UV,4\ell}}$, together with
the solution for $g'$ from Eq.  (\ref{schemetrans}) with $g =
g_{_{UV,4\ell}}$ and the fractional difference
$\Delta(g_{_{UV,4\ell}},g')$.  Since the scheme transformation
(\ref{schemetrans}) approaches the identity as $N \to \infty$, it
follows that $\lim_{N \to \infty} g' = \lim_{N \to \infty} g$ for all
$g$. In particular, as is evident from Table \ref{gprime_table}, in
the region $N \gsim 10^3$, if one sets $g=g_{_{UV,4\ell}}$, then the
corresponding value of $g'$ is close to this value. For example, for
$N=2 \times 10^3$ and $N=10^4$, the respective fractional differences
between $g_{_{UV,4\ell}}$ and the corresponding $g'$ are approximately
8 \% and 2 \% in magnitude.  Furthermore, since $b_1$ and $b_3$ are
both positive, the condition that the Jacobian $J$ should not vanish
is satisfied.  However, we find that in the region of $N \lsim N_c$,
although one can formally apply this scheme transformation, thereby
switching to a scheme in which the six-loop beta function has a UV
zero, the value of the coupling at this UV zero, $g'$, is
substantially different from $g_{_{UV,4\ell}}$. Hence, in the region
$N \lsim N_c$, this theory does not satisfy a necessary requirement
for a reliably calculable UV zero of the beta function, namely that
the values calculated in different schemes should be close to each
other.  In this region of $N$, the scheme transformation obviously
also fails to satisfy the fourth condition, C$_4$, for acceptibility
discussed above.  These results are consistent with the conclusion
that if $N \lsim N_c$, then the beta function, calculated to $O(g^4$),
does not exhibit evidence for an ultraviolet zero.


\section{Conclusions} 
\label{conclusion_section}

In this paper we have investigated the ultraviolet behavior of the
$|\vec\phi|^6_3$ theory, focusing on the question of whether, for a
given $N$, this theory exhibits robust evidence of an ultraviolet zero
in the beta function, as calculated to the six-loop (i.e., $O(g^4)$)
order.  We make use of the result for the six-loop beta function
calculated in the minimal subtraction scheme in \cite{hager}. Early
work \cite{townsend,pisarski,appelquist_heinz2} established that this
theory has a UV zero $g_{_{UV,4\ell}}$ in the four-loop beta function,
which is reliably calculable for large $N$. We find that the six-loop
beta function from \cite{hager} has a UV zero if $N > N_c$, where $N_c
\simeq 796$.  From studying the fractional difference between
$g_{_{UV,4\ell}}$ and $g_{_{UV,6\ell}}$ as a function of $N$, we
conclude that this zero in the six-loop beta function is robust for
$N$ well above $N_c$.  To study the properties of the theory for
finite $N$ further, we have analyzed the Pad\'e approximant to the
(reduced) six-loop beta function, $[1,1]_{6\ell}$. Although this
approximant does have a UV zero for a range of $N$ below $N_c$, the
value of the coupling at this UV zero, $g_{_{UV,[1,1]_{6\ell}}}$, is
not close to the value $g_{_{UV,4\ell}}$ obtained from the four-loop
beta function, so that this does not constitute evidence that the
theory actually has a reliably calculable UV zero in this range of
$N$.  Our application of a scheme transformation to the minimal
two-term beta function ('t Hooft scheme) yields the same conclusion.
Quantitatively, if one imposes the criterion that the fractional
difference between $g_{_{UV,4\ell}}$ and $g_{_{UV,6\ell}}$ should be
smaller than, say, 15 \% for the calculation of the UV zero in the
beta function to be reasonably reliable, then our results show that
this criterion is satisfied for $N \gsim 2 \times 10^3$.  Clearly,
there is some arbitrariness in this benchmark value of 15 \% for the
relative agreement of these couplings; imposing a more (less)
stringent requirement on the relative agreement of $g_{_{UV,4\ell}}$
and $g_{_{UV,6\ell}}$ would shift the estimated minimal value of $N$
for a reliable calculation to a higher (lower) value than
$2 \times 10^3$.  It is again noted that nonperturbative effects may
be important for this theory.  However, we believe that it is useful
at least to investigate the basic perturbative question of the range
in $N$ for which the beta function, calculated to the highest order to
which it is known, yields robust evidence for an ultraviolet zero. We
have addressed this question in the present paper.


\begin{acknowledgments}

  This research was partly supported by the U.S. National Science Foundation
  grant NSF PHY-22-15093. 

\end{acknowledgments}


\begin{appendix}


\section{Conversions Between Different Normalization Conventions}
\label{conversion_section}

  In the literature, several different normalization conventions have
  been used for the interaction coupling in $|\vec\phi|^6_3$ theories.
  We list some conversion relations here and remark on the
  consequences of these normalizations for the respective beta
  functions.  All of the works included here used a real $N$-component
  scalar field $\phi$ except for
  \cite{appelquist_heinz1,appelquist_heinz2}, which used a complex
  $N$-component scalar field, equivalent to a $2N$-component real
  field. Aside from numerical prefactors, there have been two general
  classes of normalization conventions.  The first class of
  normalizations involves division of the coupling by $N^2$ in the
  interaction term ${\cal L}_{\rm int.}$, while the second does not.
  We list the interaction terms below, in the notation used in the
  original papers, with superscripts added for clarity.
  Ref. \cite{townsend} by Townsend (T) used the interaction
  term
  \beq
  {\cal L}^{(\rm T)}_{\rm int.} = \frac{1}{6! N^2} \eta |\vec\phi|^6 \ ,
  \label{lagrangian_t}
  \eeq
  Ref. \cite{appelquist_heinz2} by Appelquist and Heinz (AH) used
  the interaction term (with a complex field $\vec\phi$)
  \beq
  {\cal L}^{(\rm AH)}_{\rm int.} = \frac{1}{6N^2} g
  (\vec \phi^\dagger \cdot \vec\phi)^3 \ ,
  \label{lagrangian_ah}
  \eeq
  and Ref. \cite{bmb} by Bardeen, Moshe, and Bander (BMB)
  used
  \beq
  {\cal L}^{(\rm BMB)}_{\rm int.} = \frac{1}{6N^2} \eta |\vec\phi|^6 \ .
  \label{lagrangian_bmb}
  \eeq
  Among the second class of normalizations, Ref. \cite{pisarski} by
  Pisarski (P) used 
  \beq
  {\cal L}^{(\rm P)}_{\rm int.} = \frac{\pi^2}{3} \lambda |\vec\phi|^6 \ .
  \label{lagrangian_p}
  \eeq
  while  Ref. \cite{hager} by Hager (H) used 
  \beq
  {\cal L}^{(\rm H)}_{\rm int.} = \frac{1}{6!} w |\vec\phi|^6  \ , 
  \label{lagrangian_hager}
  \eeq
and also the rescaling
  \beq
  \bar w \equiv \frac{w}{32\pi^2} \ . 
  \label{wbar}
  \eeq
  (The reader should not confuse the sextic coupling $\lambda^{(P)}$ used
  in \cite{pisarski} with the quartic coupling $\lambda$ that we have
  used in Eq. (\ref{lag}).) 
  We have employed the BMB normalization convention in our Eq. (\ref{lag})
  but with the symbol $g$ rather than $\eta$.  These couplings are
  related to each other as follows, where we use the notation in the
  original papers:
  \beqs
  && \eta^{(\rm BMB)} = \frac{1}{2}g^{(\rm AH)}
  = \frac{1}{5!} \eta^{(\rm T)} = 2\pi^2 N^2 \lambda^{(\rm P)}
  = \frac{4\pi^2 N^2}{15} \bar w^{(\rm H)} \ . \cr\cr
  &&
  \label{coupling_relations}
  \eeqs

  These different normalizations affect the definition of the
  respective beta functions. In general, consider two $|\vec\phi|^6$
  interaction couplings $c$ and $c'$ that are related to each other
  according to
  \beq
  c' = r c \ , 
  \label{cpc}
  \eeq
  where $r$ is a multiplicative factor, The corresponding beta functions
  are $\beta_c = dc/d\ln\mu$ and $\beta_{c'} = dc'/d\ln\mu$, with respective
  series expansions 
  \beq
  \beta_c = c\sum_{j=1}^\infty b_{c,j} c^j
  \label{betac}
  \eeq
  and
  \beq
  \beta_{c'} = c'\sum_{j=1}^\infty b_{c',j} c'^j \ .
  \label{betacprime}
  \eeq
  Then, since $b_{c,j}c^j = b_{c',j}c'^j = b_{c',j}(rc)^j$, it follows
  that these expansion coefficients are related according to
  \beq
  b_{c',j} = r^{-j} b_{c,j} \ . 
  \label{bcbcp}
  \eeq
  Consequently, as is evident in Eqs. (\ref{b1}), (\ref{b2}), and
  (\ref{b3}), with the T, AH, or BMB normalizations of the coupling,
  the corresponding beta function vanishes in the limit $N \to
  \infty$. Hence, if $m^2$ and $\lambda$ are tuned to zero, in this
  limit the theory is scale-invariant, and it is this scale invariance
  that was found in \cite{bmb} to be spontaneously broken if the BMB
  coupling is larger than $(4\pi)^2$. In contrast, with the
  normalization used in \cite{pisarski} and \cite{hager}, the
  respective beta functions $\beta_\lambda = d\lambda/d\ln\mu$ and
  $\beta_{\bar w} = d\bar w/d\ln \mu$ do not vanish for large $N$.
  Note that a ratio such as $b_1 b_3/b_2^2$ is invariant under
  these changes in normalizations.
  
  For reference, the conversion relations for the UV zero of the
  four-loop beta function are
  \beqs
  && \eta^{(\rm BMB)}_{_{UV,4\ell}} = 192 \ \Leftrightarrow \ 
  g^{(\rm AH)}_{_{UV,4\ell}} = 384 \ \Leftrightarrow \cr\cr
  && \lambda^{(\rm P)}_{_{UV,4\ell}} = \frac{96}{\pi^2 N^2}  \ \Leftrightarrow
            \ \bar w_{_{UV,4\ell}} = \frac{720}{\pi^2 N^2} \ . 
  \label{wuv_4loop_conversions}
  \eeqs

\end{appendix}

\bibliography{f63}

${}$

\newpage

\begin{table}
  \caption{\footnotesize{From left to right, the columns of this table list
      (i) $N$; 
      (ii) the UV zero, $g_{_{UV,4\ell}}$, of the four-loop beta function,
      $\beta_{g,4\ell}$;
      (iii) the UV zero, $g_{_{UV,6\ell}}$, of the six-loop beta
      function, $\beta_{g,6\ell}$;
      (iv) the UV zero, $g_{_{UV,[1,1]_{6\ell}}}$, of the [1,1]
      Pad\'e approximant to the reduced six-loop beta function,
      $\beta_{g,{\rm red.},6\ell}$;
      and the fractional differences, denoted for short as (v)
$\Delta_{4\ell,6\ell} \equiv \Delta(g_{_{UV,4\ell}},g_{_{UV,6\ell}})$;
      (vi) $\Delta_{4\ell,[1,1]_{6\ell}} \equiv
        \Delta(g_{_{UV,4\ell}},g_{_{UV,[1,1]_{6\ell}}})$; and
     (vii) $\Delta_{6\ell,[1,1]_{6\ell}}  \equiv
        \Delta(g_{_{UV,6\ell}},g_{_{UV,[1,1]_{6\ell}}})$. The last
        row lists the limiting values as $N \to \infty$. We use 
        the standard notation $-0.331$e-2 for $-(0.331 \times 10^{-2})$, etc.
        The symbol ``n'' means that the entry is unphysical or not relevant.}}
  \begin{center}
  \begin{adjustbox}{width=0.5\textwidth}
\begin{tabular}{|c|c|c|c|c|c|c|} \hline\hline
  $N$ & $g_{_{UV,4\ell}}$ & $g_{_{UV,6\ell}}$ &
  $g_{_{UV,[1,1]_{6\ell}}}$ & $\Delta_{4\ell,6\ell}$ &
  $\Delta_{4\ell,[1,1]_{6\ell}}$ & $\Delta_{6\ell,[1,1]_{6\ell}}$ 
\\ \hline 
1     & 0.2356  & n       & n       & n        & n        &  n \\
10    & 12.21   & n       & n       & n        & n        &  n \\
100   & 108.09  & n       & n       & n        & n        &  n \\
300   & 154.71  & n       & 378.17  & n        & 1.44    &  n \\
900   & 178.05  & 268.07  & 229.16  & 0.506   & 0.287   & $-0.145$  \\
1.0e3 & 179.37  & 249.49  & 224.79  & 0.391   & 0.253   & $-0.0990$ \\
2.0e3 & 185.505 & 210.34  & 207.07  & 0.134   & 0.116   & $-0.0156$ \\
4.0e3 & 188.71  & 199.90  & 199.24  & 0.0593  & 0.0558  & $-0.331$e-2 \\
1.0e4 & 190.67  & 194.93  & 194.83  & 0.0223  & 0.0218  & $-0.513$e-3 \\
$\infty$ &192   & 192     & 192     & 0       & 0       & 0          \\
\hline\hline
\end{tabular}
\end{adjustbox}
\end{center}
\label{guv_table}
\end{table}
%


\normalsize

\begin{table}
  \caption{\footnotesize{From left to right, the columns of this table
      list (i) $N$; (ii) the UV zero, $g_{_{UV,4\ell}}$, of the
      four-loop beta function, $\beta_{g,4\ell}$; (iii)
      $g'$, the value of $g$ in the transformed scheme with
      $b_3'=0$ obtained via the solution of Eq.  (\ref{schemetrans}) with
      $g$ set equal to $g_{_{UV,4\ell}}$; and (iv) the fractional
      difference between these values, denoted for short as
      $\Delta_{\rm tran.} \equiv \Delta(g_{_{UV,4\ell}},g')$. The
      last row lists the limiting values as $N \to \infty$. We use
      the standard notation 1.0e3 for $1.0 \times 10^3$, etc.}}
\begin{center}
\begin{tabular}{|c|c|c|c|} \hline\hline
  $N$ & $g_{_{UV,4\ell}}$ & $g'$ & $\Delta_{\rm tran.}$   
\\ \hline
100      & 108.09     & 69.901     & $-0.353$   \\ 
300      & 154.71     & 116.09     & $-0.250$   \\
900      & 178.05     & 152.90     & $-0.141$   \\ 
1.0e3    & 179.37     & 155.67     & $-0.132$   \\
2.0e3    & 185.505    & 170.50     & $-0.0809$  \\
4.0e3    & 188.71     & 180.04     & $-0.0459$  \\
1.0e4    & 190.67     & 186.84     & $-0.0201$  \\
$\infty$ & 192        & 192        & 0          \\
\hline\hline
\end{tabular}
\end{center}
\label{gprime_table}
\end{table}

\end{document}